\documentclass[reprint,amsmath,amssymb,aps]{revtex4-1}
\usepackage{graphicx,dcolumn,bm,amssymb,amstext,amsmath,cases,float,setspace,mathtools,amsthm,physymb}
\pdfoutput=1

\usepackage[none]{hyphenat}

\begin{document}
\title{Discrete nature of thermodynamics in confined ideal Fermi gases}
\thanks{NOTICE: This is the author’s version of a work that was accepted for publication in Physics Letters A. Changes resulting from the publishing process, such as peer review, editing, corrections, structural formatting, and other quality control mechanisms may not be reflected in this document. Changes may have been made to this work since it was submitted for publication. A definitive version was subsequently published in Phys. Lett. A, [378, 30-31, 6/13/2014] DOI: 10.1016/j.physleta.2014.05.044}%
\author{Alhun Aydin}
\author{Altug Sisman}%
 \email{Corresponding author, sismanal@itu.edu.tr}
\affiliation{Istanbul Technical University, Energy Institute, 34469, Istanbul, Turkey}
\date{\today}
\begin{abstract}
Intrinsic discrete nature in thermodynamic properties of Fermi gases appears under strongly confined and degenerate conditions. For a rectangular confinement domain, thermodynamic properties of an ideal Fermi gas are expressed in their exact summation forms. For 1D, 2D and 3D nano domains, variations of both number of particles and internal energy per particle with chemical potential are examined. It is shown that their relation with chemical potential exhibits a discrete nature which allows them to take only some definite values. Furthermore, quasi-irregular oscillatory-like sharp peaks are observed in heat capacity. New nano devices can be developed based on these behaviors.
\end{abstract}
\maketitle
\section{Introduction}
Nano scale thermodynamics may be regarded as a relatively new research area. Nevertheless, considering leap forwards in nanoscience and nanotechnologies in recent years, it is fair to say that understanding thermodynamics of nano systems became an inevitable necessity [1-13]. When dealing with thermodynamics of nano systems, it is crucial to understand in which ways they differ from macro systems.

Thermodynamic state functions are written as infinite summations, in their exact forms. In classical thermodynamics, sums are replaced by integrals, under continuum approximation \cite{14}. However, this approximation is no longer valid at nano scale and thereby, summations and integrals give considerably different results. Therefore, either summation forms of state functions have to be calculated numerically, or derived analytically by using more precise methods like Poisson summation formula etc., in order to examine thermodynamic behaviors of systems at nano scale.

To evaluate summations analytically, Poisson summation formula (PSF) for even functions is used and it is given as follows
\begin{equation}
\sum^{\infty}_{i=1} f(i)=\int^{\infty}_{0} f(i)di - \frac{f(0)}{2}+2\sum^{\infty}_{s=1}\int^{\infty}_{0} f(i)cos(2\pi si)di
\end{equation}

First term of PSF is the conventional integral term and the second term represents the zero correction which is considered as quantum size effects in literature. For Maxwell-Boltzmann gases, the second term is responsible for quantum boundary layer, anisotropic gas pressure, non-additivity of global thermodynamic properties, quantum forces and thermosize effects which are similar to thermoelectric effects [5, 6, 8, 9, 11-13]. Third term of the PSF is the discrete correction term to the integral representation of summations. While the contributions of both the second and third terms are negligible in macro scale, they become substantial in strongly confined nano structures. The contribution of third term becomes crucial only if at least one of the domain sizes is smaller than the de Broglie wavelength of particles. When analytical solution for the third term is not obtainable, it is preferred to calculate the summation itself numerically instead of calculating the whole terms of PSF.

In this paper, by considering the exact forms of thermodynamic state functions based on infinite sums, it is shown that the intrinsic discrete nature appears in thermodynamic quantities such as chemical potential, internal energy per particle and heat capacity per particle at constant volume for strongly confined and degenerate ideal Fermi gases. Besides the numerical results, some analytical expressions, representing the discrete nature of chemical potential and internal energy per particle, are derived and gaps between their discrete values are analytically expressed. Furthermore, oscillatory nature of heat capacity varying with particle number and domain size is observed. These behaviors are in agreement with the results of \cite{15, 16} in which chemical potential and specific heat have been just numerically calculated for ultra-cold Fermi gases confined in harmonic and quartic traps. Surface density dependency of heat capacity for 2D Fermi gases has been considered as the manifestation of QSE in \cite{17}. In addition, temperature dependence of specific heat capacity of non-interacting fermions with multifractal energy spectra has been numerically analyzed in literature \cite{18, 19}. On the other hand, in this study, there is no restriction for temperature and analytical expressions representing the discrete nature of thermodynamic properties are given. Variation of heat capacity with domain size is also discussed as a manifestation of quantum size effects. Furthermore, oscillations and sharp peaks in heat capacity are explained by defining the discrete Fermi point, line and surface.

Discrete and oscillatory behaviors do not appear in Bose gases and they disappear even in Fermi gas at high temperature limit or for large domain sizes in comparison with the de Broglie wavelength of particles. In other words, these behaviors appear only when quantum degeneracy and size effects are strong enough to make the Pauli exclusion principle and quantum character of particles dominant in behavior of the system. 
\section{Intrinsic Discrete Nature of Thermodynamic Properties in An Ideal Fermi Gas Confined at Nano Scale}
Considering mono particles (like monatomic gas, electron gas etc.) confined in a rectangular domain with dimensions $L_1, L_2$ and $L_3$, energy eigenvalues of particles from Schr\"{o}dinger equation are
\begin{equation}
\begin{multlined}
\varepsilon=\frac{h^2}{8m}\sum^{3}_{n=1}\left(\frac{i_n}{L_n}\right)^2, \mbox{   with   } {i_n}=1,2,3,...,
\end{multlined}
\end{equation}
where $h$ is the Planck's constant, $m$ is the mass of particle and $n=1,2,3$ representing the spatial directions. In order to discuss confinement magnitudes of different domains, confinement parameters can be defined as $\alpha_n=L_c/L_n$ where $L_c$ is a length scale based on de Broglie wavelength of particles that is expressed as $L_c=h/\sqrt{8mk_b T}$, where $k_B$ is the Boltzmann's constant and $T$ is the temperature of the gas.
Then, dimensionless energy eigenvalues in $\alpha$-notation is
\begin{equation}
\frac{\varepsilon}{k_B T}=\tilde{\varepsilon}=[(\alpha_1 i_1)^2 + (\alpha_2 i_2)^2 + (\alpha_3 i_3)^2]
\end{equation}
Note that, confinement parameter $\alpha$ is inversely proportional to the domain size, so as the name implies, increasing $\alpha$ means increasing confinement and when $\alpha$ becomes greater than unity, domain size is smaller than $L_c$.

By neglecting spin degrees of freedom $g_s$ (which is just a factor of 2 for a free electron gas or $He_3$), Fermi-Dirac distribution function is
\begin{equation}
f=\frac{1}{\exp(\tilde{\varepsilon}-\Lambda)+1}
\end{equation}
where $\Lambda$ is dimensionless chemical potential, defined as $\Lambda=\mu /k_B T$.
Summation of the distribution function over momentum states in all directions gives the number of particles in the system
\begin{equation}
N=\sum^{\infty}_{i_n=1}\frac{1}{\exp(\tilde{\varepsilon}-\Lambda)+1}
\end{equation}
where $i_n=i_1,i_2,i_3$, so it is a triple sum.

In this letter, we consider an ideal Fermi gas confined in three different rectangular domains. Temperature is kept constant in all cases considered here. In the first case, domain is strongly confined in two directions (domain sizes are much smaller than $L_c$ in those directions) and relatively weakly confined in the other direction. In strong confinements, values of momentum component of that direction are always equal to the value of the ground state. Thus, no summations have to be done over that momentum components. In relatively weakly confined directions, however, particles can become excited so that value of momentum components in that directions can take the values of excited states and summations have to be done.
\subsection{1D Fermi Gas}
Although thermodynamic quantities are calculated based on infinite sums, after Fermi level ($i_F=\sqrt{\Lambda}/\alpha$), contributions to the summation rapidly decrease and it is quite sufficient to sum up to $i_{max}=2i_F$. After $i_{max}$, contributions become completely negligible. Moreover, for directions with $\alpha$ values much larger than unity, $i_F$ goes to zero for a limited range of $\Lambda$ and this completely removes the summation process for that direction.

Let's consider an anisometric domain with confinement parameters $\alpha_1=1$, $\alpha_2=40$ and $\alpha_3=40$. Note that, $40$ is a realistic value for $\alpha$, since for an electron confined in a single atom carbon layer (graphene) with $0.3$ nm thickness, $\alpha=43$ at $20$K \cite{20}. In this system, we can replace triple summations by a single sum as long as $\Lambda<\Lambda_1=(\alpha_1)^2+(\alpha_2)^2+(2\alpha_3)^2$. After the value of $\Lambda_1$, particles can occupy the excited states of momentum components even in strongly confined directions and they make considerable contributions to the summation. As a result, for a definite range of $\Lambda$, triple sum can be represented by a single sum, since for strongly confined directions only ground state of momentum components contributes. 

In case of $\alpha_1 << 1$, gas is still 1D Fermi gas since $\left\{\alpha_2, \alpha_3\right\}>>1$ and it is enough to use first two terms of PSF to calculate the summation. In that case, total number of particles, Eq. (5), can be approximated for a 1D Fermi gas as
\begin{subequations}
\begin{align}
N_{1D} & \approx -\frac{\sqrt{\pi}}{2\alpha_1}Li_{1/2}[-\exp({\Lambda^\prime})]+\frac{1}{2}Li_0[-\exp(\Lambda^\prime)] \\
       & \approx \frac{\sqrt{\Lambda^\prime}}{\alpha_1}-\frac{1}{2}
\end{align}
\end{subequations}
where $\Lambda^\prime=\Lambda-\alpha_{2}^2-\alpha_{3}^2$ and $Li$ denotes the polylogarithm function. Eq. (6b) is obtained by using the asymptotic approximations of polylogarithm functions for $\Lambda>>1$ (strongly degenerate Fermi gas). On the other hand, if $\alpha_1$ is not much less than unity, namely in nano scale, the whole terms of PSF have to be considered to evaluate the summation properly and analytically. Therefore, by using the whole terms of PSF and asymptotic forms of polylogarithm functions, after some mathematical operations, 1D number of particles is obtained as
\begin{equation}
N_{1D}^{PSF} \cong \frac{\sqrt{\Lambda^\prime}}{\alpha_1}-\frac{1}{2}+\frac{1}{\pi}\arctan\left[\cot\left(\frac{\pi\sqrt{\Lambda^\prime}}{\alpha_1}\right)\right]
\end{equation}
where PSF superscript indicates that the whole terms of PSF are used to obtain the expression. Variation of number of particles with dimensionless chemical potential $\Lambda$ for anisometric 1D Fermi gas is given in Figure 1.

\begin{figure}[H]
\centering
\includegraphics[width=0.47\textwidth]{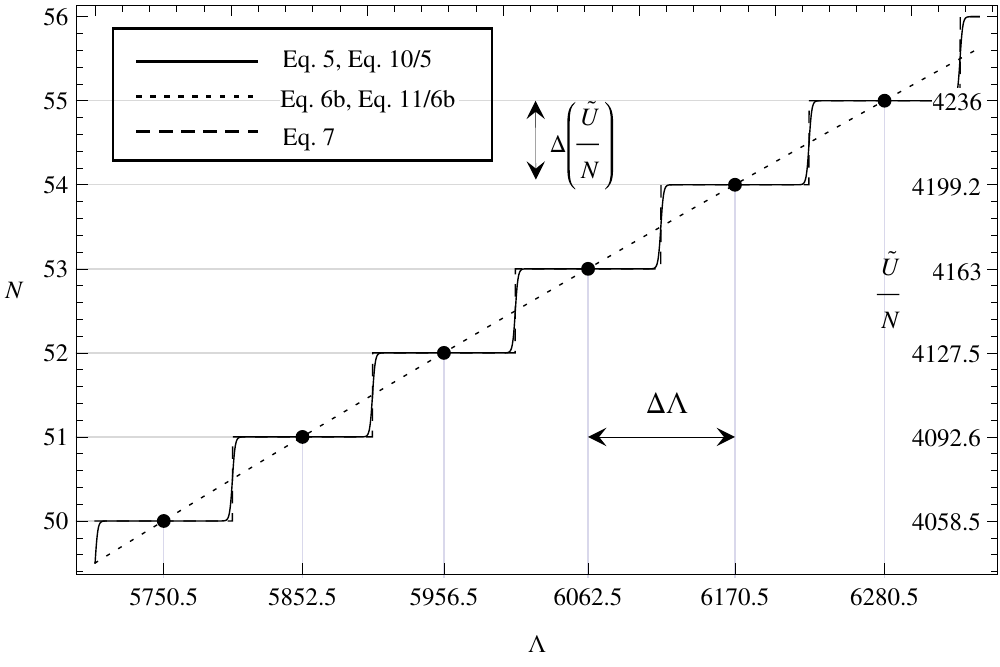}
\caption{Number of particles (left axis) and dimensionless internal energy per particle (right axis) varying with dimensionless chemical potential for 1D Fermi gas with $\alpha_1=1, \alpha_2=40, \alpha_3=40$.}
\label{fig:pic1}
\end{figure}

In Fig. 1, the variation of number of particles and dimensionless internal energy per particle with dimensionless chemical potential is given in left and right axes respectively. Considering the left axis first; the solid curve represents Eq. (5), that is the exact expression for number of particles confined in a 1D nano domain. Bold dots indicate the $\Lambda$ values which correspond to the integer particle numbers and they are solved numerically from Eq. (5). Dashed curve represents Eq. (7), which gives almost the same result of Eq. (5) for 1D, except the sharpness of the edges of steps. Therefore, Eq. (7) correctly predicts that chemical potential is a discrete quantity as long as number of particles is integer and this discreteness of chemical potential becomes important in a degenerate Fermi gas confined in a nano domain. Dotted curve is obtained from Eq. (6b) and it passes exactly from the points where the particle number is integer even though it cannot represent the true discrete relation between number of particles and chemical potential.

For 1D Fermi gases that are strongly confined in second and third directions in a rectangular domain, a formula that gives the discrete $\Lambda$ values corresponding to the integer number of particles can be defined. Either the third term of Eq. (7) is equalized to zero to find its roots, or more easily $\Lambda$ can analytically be solved from the first two terms of Eq. (7) since third term is exactly zero when number of particles is integer. Then, for integer $N>>1$
\begin{equation}
\Lambda^\prime=\alpha_{1}^2\left(N+\frac{1}{2}\right)^2
\end{equation}
As it is seen, Eq. (8) gives the possibility to write thermodynamic properties of 1D Fermi gases in terms of particle number instead of chemical potential. Besides, from Eq. (8), expression for dimensionless chemical potential intervals can easily be obtained
\begin{equation}
\Delta\Lambda^\prime=2\alpha_{1}^2(N+1)=2\alpha_{1}\sqrt{\Lambda^\prime}+\alpha_{1}^2
\end{equation}

The other thermodynamic quantity that represents stepwise nature is the internal energy. Dimensionless internal energy is written in its exact form as
\begin{equation}
\frac{U}{k_B T}=\widetilde{U}=\sum^{\infty}_{i_n=1}\frac{\tilde{\varepsilon}}{\exp(\tilde{\varepsilon}-\Lambda)+1}
\end{equation}

Using the first two terms of PSF and asymptotic approximations of polylogarithm functions, dimensionless internal energy for 1D Fermi gases with integer number of particles is obtained as
\begin{equation}
\widetilde{U}_{1D}=\frac{(\Lambda^\prime)^{3/2}}{3\alpha_1}+\left(\frac{\sqrt{\Lambda^\prime}}{\alpha_1}-\frac{1}{2}\right)(\alpha_{2}^2+\alpha_{3}^2)
\end{equation}

As long as particle number is integer, there are no longer continuous $\Lambda$ values, and since all specific (per particle) thermodynamic state functions depend on $\Lambda$, specific internal energy and heat capacity have also discrete values corresponding to the integer particle number.

Considering the right axis of Fig. 1, discreteness of internal energy per particle is shown for 1D Fermi gas for particle numbers from 50 to 55 where the functional behavior is exactly the same as the $N-\Lambda$ relation. This time, solid curve is obtained by dividing Eq. (10) to Eq. (5). Dotted line represents the result of Eq. (11) divided by Eq. (6b) and it matches with exact solution for integer $N$. If $N$ is large enough, even for the unit value of $\alpha$, $\Lambda^\prime$ becomes very large, so that by using Eq. (8) and Eq. (11) dimensionless internal energy is written in terms of $N$ and $\alpha$ as
\begin{equation}
\widetilde{U}_{1D}\cong \frac{\alpha_{1}^2}{3}\left(N+\frac{1}{2}\right)^3+N(\alpha_{2}^2+\alpha_{3}^2)
\end{equation}

Using Eq. (12), dimensionless internal energy interval per particle for 1D Fermi gas, can be derived as
\begin{equation}
\Delta\left(\frac{\widetilde{U}}{N}\right)\cong\frac{2}{3}\alpha_{1}^2 \left(N+\frac{5}{4}\right)
\end{equation}

From Eq. (13), it can be deduced that when $\alpha$ goes to zero (as in macro systems), discrete nature and stepwise behavior practically disappears although Eq. (10) preserves its intrinsic discrete nature. Conversely, for large $\alpha$ values it become apparent. In addition, discreteness in $N-\Lambda$ or $\widetilde{U}-\Lambda$ relations depend on confinement parameter $\alpha$ instead of temperature alone. So, even at $T=0$K, steepness of the function in Fig. 1 does not get sharper as long as $\alpha$ is kept constant. On the other hand, both dimensionless chemical potential and internal energy intervals change with the value of $\alpha$ which depends on temperature, Eq. (9) and Eq. (13). When temperature goes to zero, $\alpha$ values and so the discreteness interval of $\Lambda$ go to infinity.
\subsubsection{Fermi Point and Zero Heat Capacity in confined 1D Fermi gas}
Now let's examine the heat capacity of 1D Fermi gas at nano scale. Dimensionless heat capacity at constant volume is written in its exact form as
\begin{equation}
\frac{C_V}{k_B}=\widetilde{C}_V=\sum_{i_n}(\tilde{\varepsilon})^2f(1-f)-\frac{[\sum_{i_n}\tilde{\varepsilon}f(1-f)]^2}{\sum_{i_n}f(1-f)}
\end{equation}
Derivation of heat capacity is done by differentiating internal energy $U$ with respect to temperature $T$. In the derivation of exact expression of heat capacity, to be able to interchange derivative operator with sum operator, uniform convergency of the function is checked by comparing the analytical derivative with numerical one. It is seen that results perfectly match with each other.

As it's seen from Eq. (14), heat capacity expression contains the variance of Fermi-Dirac distribution function $Var(f)=f(1-f)$, which is equal to the derivative of distribution function with respect to $\Lambda$, $df / d\Lambda=f(1-f)$. It is not possible to obtain an analytical expression for heat capacity by using the whole terms of PSF even for large values of $\Lambda$. For this reason, heat capacity is numerically calculated based on its summation form given by Eq. (14).

In degenerate Fermi gas ($\Lambda>>1$), electronic contribution to heat capacity only comes from the momentum states around Fermi point, Fermi line and Fermi surface for 1D, 2D and 3D Fermi gases respectively. The state corresponding to the Fermi point is the one for which the distribution function is equal to $1/2$ and the variance reaches its maximum value of $1/4$, as it is seen in Fig 2. Full band width at half maximum (FWHM) of variance for 1D distribution function can be found by equating $f(1-f)$ to $1/8$
\begin{equation}
\delta_{1D}=\frac{\mathrm{arcosh}(17)}{2\alpha_1 \sqrt{\Lambda^\prime}}
\end{equation}

\begin{figure}[t]
\centering
\includegraphics[width=0.44\textwidth]{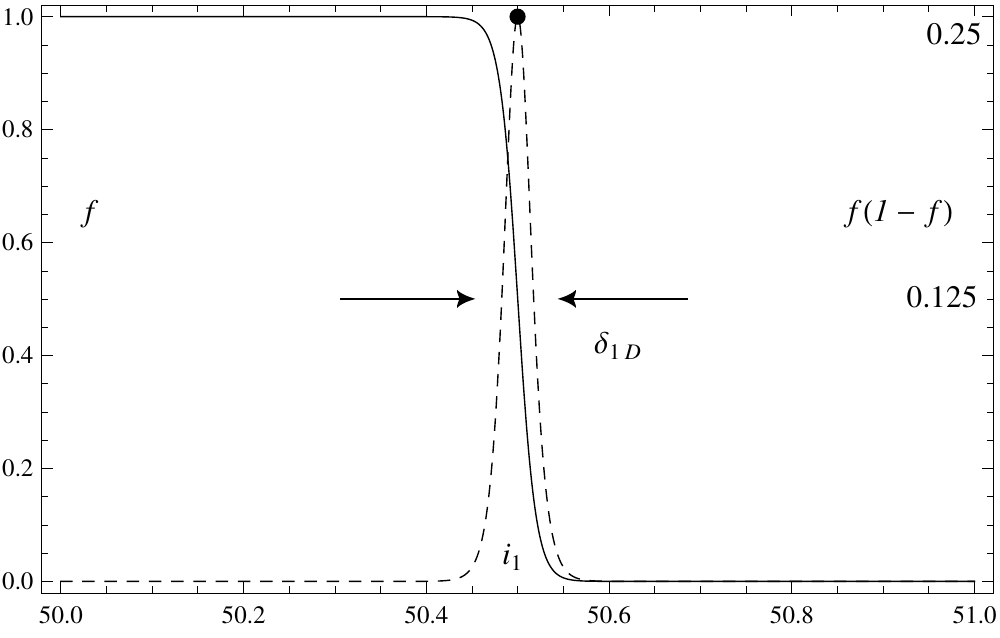}
\caption{Distribution function (left axis) and its variance (right axis) around Fermi point of 1D Fermi gas with $\alpha_1=1, \alpha_2=40, \alpha_3=40$ and $N=50$.}
\label{fig:pic1}
\end{figure}

It is easy to see from Eq. (15) that if the confinement ($\alpha$) or degeneracy ($\Lambda$) increases, FWHM decreases, so the peak of variance becomes too sharp that integer values of momentum state variable, $i_1$, correspond to the vanished values of the variance function, $f(1-f)$. In other words, Fermi point corresponds to the half integer value of $i_1$, see Fig. 2, which is not a valid value for a momentum state. In that case, contributions to heat capacity from integer values of $i_1$ are almost zero. As a result, heat capacity of 1D Fermi gas is nearly zero. When 0D system is considered, it is clear from Eq. (14) that summations die out and $C_V$ becomes exactly zero since there is only ground state in 0D system. Hence it is worth to notice that contributions from momentum states to $C_V$ always come from $D-1$ dimensional momentum space. For instance, in degenerate 2D and 3D Fermi gases, contributions come from the states on Fermi line and Fermi surface respectively.
\subsection{2D and 3D Fermi Gases}
Two different domains having 2D and 3D momentum spaces are considered in this section. Their $\alpha$ values are $\alpha_1=3$, $\alpha_2=3$, $\alpha_3=40$ and $\alpha_1=3$, $\alpha_2=3$, $\alpha_3=3$ respectively. These two cases are considered under the same section due to their similar nature. In 2D case, domain is relatively weakly confined in first and second directions whereas strongly confined in third one. For this kind of domain, there is no need to make summation over the states of momentum component in third direction since the only contribution comes from its ground state. By using Eq. (5) and the first two terms of PSF, an analytical expression for number of particles can be obtained as 
\begin{equation}
N_{2D} \approx \frac{\pi\Lambda^{\prime\prime}}{4\alpha_1 \alpha_2}-\frac{\sqrt{\Lambda^{\prime\prime}}}{2}\left(\frac{1}{\alpha_1}+\frac{1}{\alpha_2}\right)+\frac{1}{4}
\end{equation}
where $\Lambda^{\prime\prime}=\Lambda-\alpha_{3}^2$. By using the first two terms of PSF, internal energy expression for 2D Fermi gas is obtained as
\begin{equation}
\widetilde{U}_{2D} \approx \frac{\pi{\Lambda^{\prime\prime}}^2}{8\alpha_1 \alpha_2}-\frac{{\Lambda^{\prime\prime}}^{3/2}}{6}\left(\frac{1}{\alpha_1}+\frac{1}{\alpha_2}\right)+\alpha_{3}^2 N_{2D}
\end{equation}

\begin{figure}[t]
\centering
\includegraphics[width=0.45\textwidth]{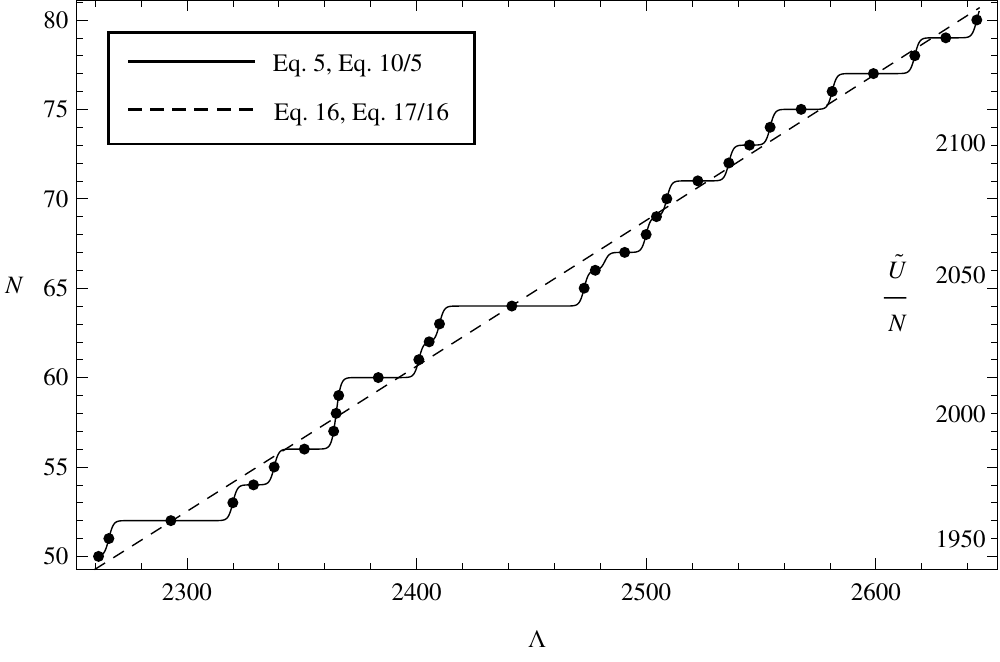}
\caption{Number of particles (left axis) and dimensionless internal energy per particle (right axis) varying with dimensionless chemical potential for 2D Fermi gas with $\alpha_1=3, \alpha_2=3, \alpha_3=40$.}
\label{fig:pic3}
\end{figure}

As it's seen from Fig. 3, quasi-irregular stepwise relation of number of particles and internal energy changing with chemical potential is observed in 2D Fermi gases, unlike the regular stepwise nature in 1D. In Fig. 3, for the left axis, Eq. (16) represented by dashed line does not perfectly matches with the exact solution based on Eq. (5) represented by solid line. On the right axis of Fig. 3, quasi-irregular stepwise behavior is observed for dimensionless internal energy per particle. Solid curve represents Eq. (10) divided by Eq. (5) and the dashed curve is the result of Eq. (17) divided by Eq. (16). Bold points mark the values where the particle number is integer.

Similar quasi-irregular discrete nature of 2D appears also in 3D Fermi gas with $\alpha_1=\alpha_2=\alpha_3=3$. In a similar way used for 2D Fermi gas, approximate analytical expressions for number of particles and internal energy of 3D Fermi gas are obtained as follows 
\begin{equation}
\begin{multlined}
N_{3D} \approx \frac{\pi\Lambda^{3/2}}{6\alpha_1 \alpha_2 \alpha_3}-\frac{\pi\Lambda}{8}\left(\frac{1}{\alpha_1 \alpha_2}+\frac{1}{\alpha_2 \alpha_3}+\frac{1}{\alpha_1 \alpha_3}\right) \\
+\frac{\sqrt{\Lambda}}{4}\left(\frac{1}{\alpha_1}+\frac{1}{\alpha_2}+\frac{1}{\alpha_3}\right)-\frac{1}{8}
\end{multlined}
\end{equation}

\begin{equation}
\begin{multlined}
\widetilde{U}_{3D} \approx \frac{\pi \Lambda^{5/2}}{10 \alpha_1 \alpha_2 \alpha_3}-\frac{\pi \Lambda^2}{16}\left(\frac{1}{\alpha_1 \alpha_2}+\frac{1}{\alpha_2 \alpha_3}+\frac{1}{\alpha_1 \alpha_3}\right) \\
+\frac{\Lambda^{3/2}}{12}\left(\frac{1}{\alpha_1}+\frac{1}{\alpha_2}+\frac{1}{\alpha_3}\right)-\frac{1}{8}
\end{multlined}
\end{equation}

As it is seen in Fig. 4, Eqs. (18) and (19) represent the trends of exact solutions quite well. Solid curve in Fig. 4 represent the results from the exact expressions based on Eqs. (5) and (10) for number of particles and dimensionless internal energy per particle respectively. Bold points indicate the values of $\Lambda$ corresponding to the integer number of particles, $N=\left\{50,...,80\right\}$. Dashed line represent the approximate analytical solutions, Eq. (18) and Eq. (19)/(18), based on the first two terms of PSF.

\begin{figure}[t]
\centering
\includegraphics[width=0.45\textwidth]{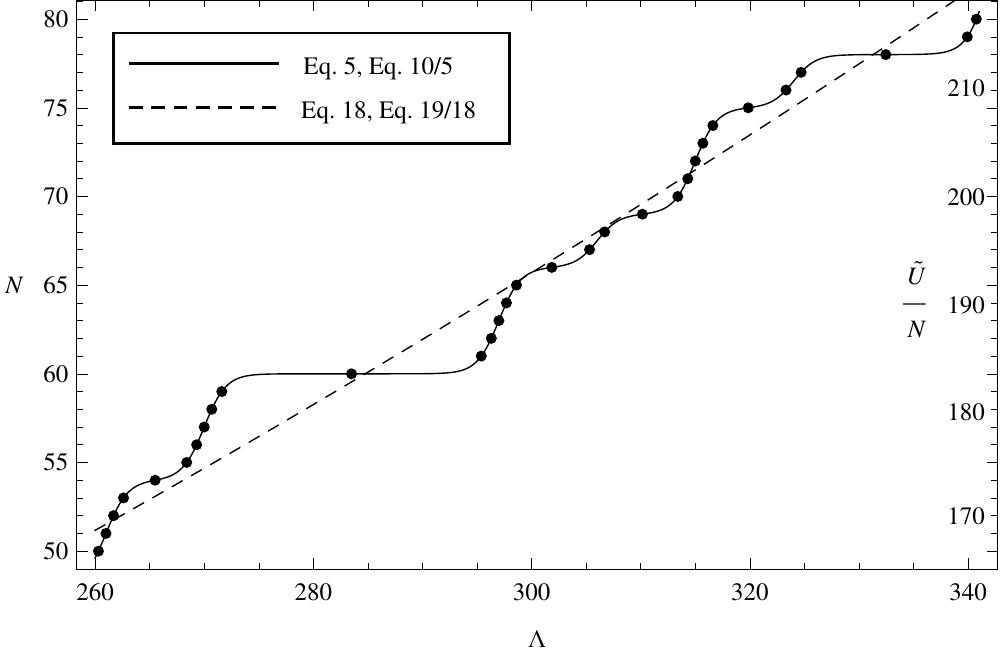}
\caption{Number of particles (left axis) and dimensionless internal energy per particle (right axis) varying with dimensionless chemical potential for 3D Fermi gas with $\alpha_1=3, \alpha_2=3, \alpha_3=3$.}
\label{fig:pic3}
\end{figure}

Interestingly, behaviors of 2D and 3D Fermi gases are quite different than that of 1D Fermi gases. To understand the reason, it is necessary to examine the nature of summations in detail. All thermodynamic properties are represented by summations over momentum state variables $\left\{i_1,i_2,i_3\right\}$, which constitute a state matrix. Each element of the state matrix represented by $\left\{i_1,i_2,i_3\right\}$, contributes to the summation differently. This state matrix can be decomposed into diagonal $\left\{i_1=i_2=i_3\right\}$ and non-diagonal matrices, except for 1D Fermi gas in which state matrix becomes a state vector. Quasi-irregular behaviors observed in 2D and 3D Fermi gases are the consequences of the contribution of the elements of non-diagonal state matrix. Unlike diagonal matrix, in non-diagonal matrix, there are mostly more than one element (momentum state) that make the same contribution to the summation, which is commonly called as degeneracy of energy levels. Since the degeneracy of energy levels changes due to the different combinatoric nature of double or triple summations, quasi-irregular patterns in Figs. 3 and 4 are different for 2D and 3D Fermi gases. On the other hand, regular stepwise behaviors instead of quasi-irregular one appear in 1D Fermi gas since there is no such degeneracy in energy levels in 1D case.
\subsubsection{Discrete Fermi Line and Heat Capacity Oscillations in confined 2D Fermi gas}
In 2D Fermi gas, contribution to heat capacity comes from the particles on the Fermi line. In other words, contributions from the particles outside of the Fermi line are totally negligible, in similar to outside of Fermi point in 1D Fermi gas. Therefore, it is enough to make summations just over the states on Fermi line to calculate the heat capacity of a 2D Fermi gas. To identify the Fermi line, it is useful to analytically define Round, Ceiling and Floor functions. Analytical expression of round of $x$ is simply given by
\begin{equation}
\mbox{Round}(x)=x+\frac{1}{\pi}\arctan\left[\cot\left(\pi\left(x+\frac{1}{2}\right)\right)\right]
\end{equation}

One can then easily obtain the Ceiling and the Floor functions by adding $1/2$ to and subtracting $1/2$ from $x$ in Eq. (20) respectively. Hence, the states on Fermi line are represented by $i_1$ from 1 to $i_{1max}$ and $i_2$ from $i_{2min}$ to $i_{2max}$ where $i_{1max}=\sqrt{\Lambda^{\prime\prime}}/\alpha_1 + 1/2$, $i_{2min}=\mbox{Ceiling}(\sqrt{\Lambda^{\prime\prime}-[\alpha_1(i_1+1/2)]^2}/\alpha_2 - 1/2)$ and $i_{2max}=\mbox{Floor}(\sqrt{\Lambda^{\prime\prime}-[\alpha_1(i_1-1/2)]^2}/\alpha_2 + 1/2)$. Idealized (continuous) and exact (discrete) Fermi lines are shown in Figure 5 by dotted curve and solid stepwise function respectively. Contribution to heat capacity comes only from the states corresponding to the couples of integer $i_1$ and $i_2$, which are represented by bold dots in Fig. 5. Two dashed curves enclosing the discrete Fermi line are $\pm 1/2$ neighborhoods of the dotted (idealized) Fermi line. These two dashed curves define a Fermi shell originated from the Heisenberg uncertainty principle. Exact Fermi line always stays within Fermi shell. The closer a state inside the Fermi shell to the idealized Fermi line, the larger its contribution to heat capacity and vice versa.

\begin{figure}[t]
\centering
\includegraphics[width=0.34\textwidth]{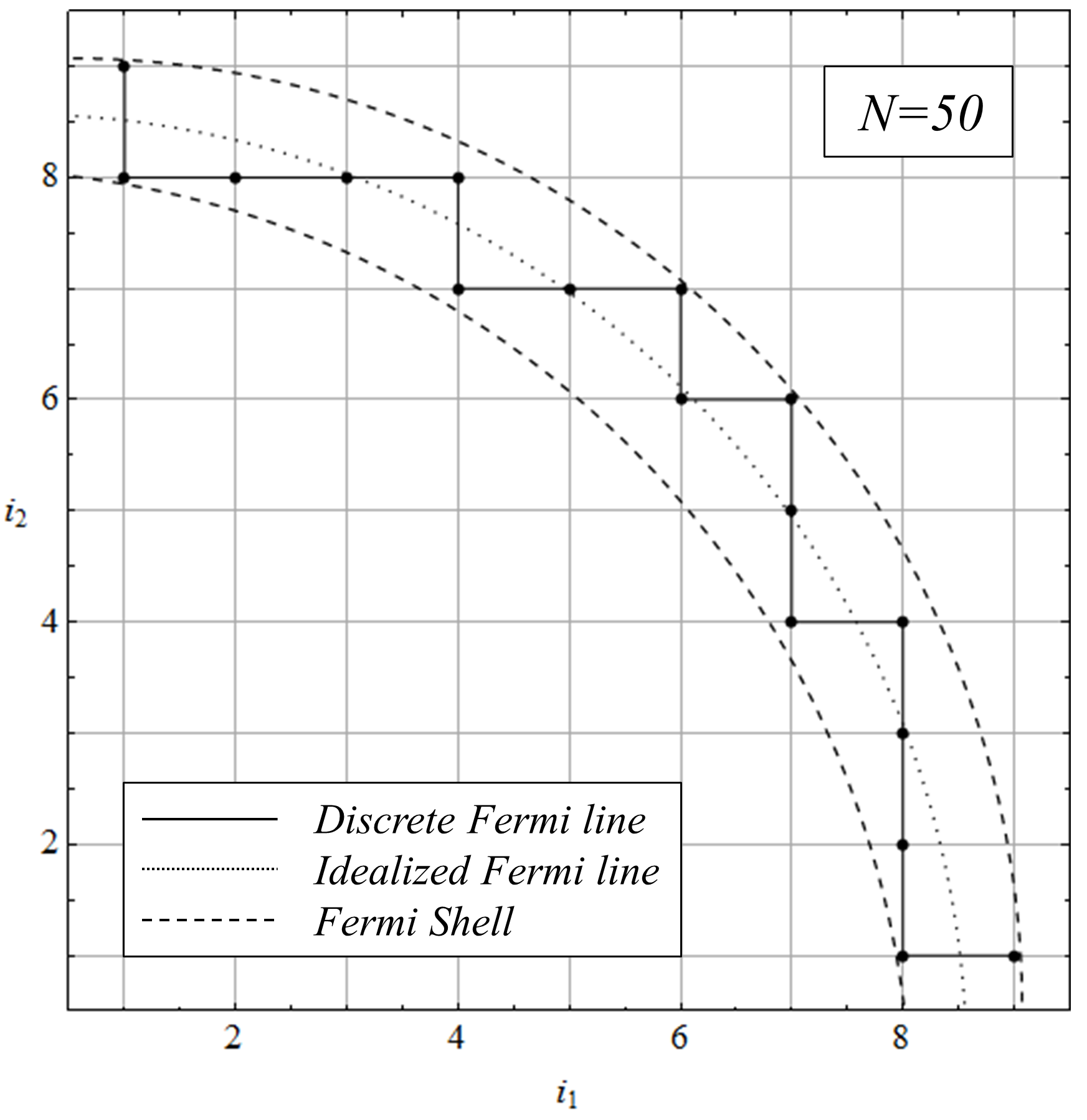}
\caption{Idealized and discrete Fermi lines and Fermi shell for 2D Fermi gas with $\alpha_1=3, \alpha_2=3, \alpha_3=40$ and $N=50$.}
\label{fig:pic3}
\end{figure}

In Figure 6, peakwise nature is seen in the variation of dimensionless specific heat capacity of 2D Fermi gas with number of particles.

\begin{figure}[b]
\centering
\includegraphics[width=0.47\textwidth]{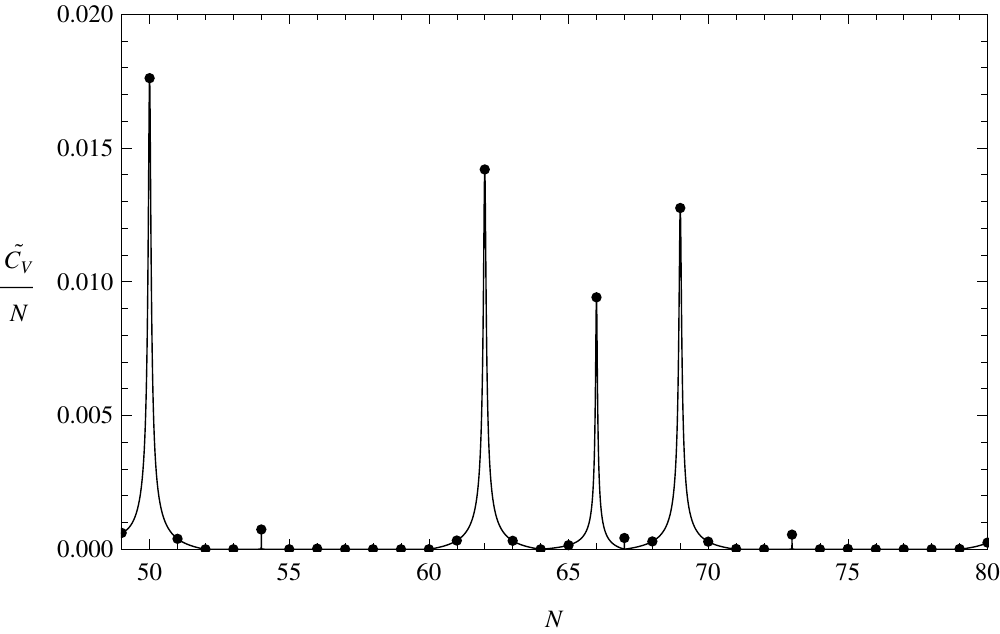}
\caption{Dimensionless heat capacity per particle vs number of particles for 2D Fermi gas with $\alpha_1=3, \alpha_2=3, \alpha_3=40$.}
\label{fig:pic3}
\end{figure}

When $N$ is increased, Fermi line and consequently the Fermi shell extends while its thickness remains constant. Since the combination of the proximities of the states in Fermi shell to the idealized Fermi line changes, the magnitude of their contributions also changes and causes the peaks in heat capacity for certain number of particles. For example, in Fig. 6, there is a peak for 50 particles and it goes almost down to zero for 52 particles. As it is seen from Fig. 5, there are several states (bold points) which are close to the idealized Fermi line for 50 particles, so their contribution are large enough to generate a peak in heat capacity. Conversely, in Figure 7, for 52 particles, states in Fermi shell are mostly far from the idealized Fermi line, so heat capacity goes down to almost zero.

\begin{figure}[t]
\centering
\includegraphics[width=0.34\textwidth]{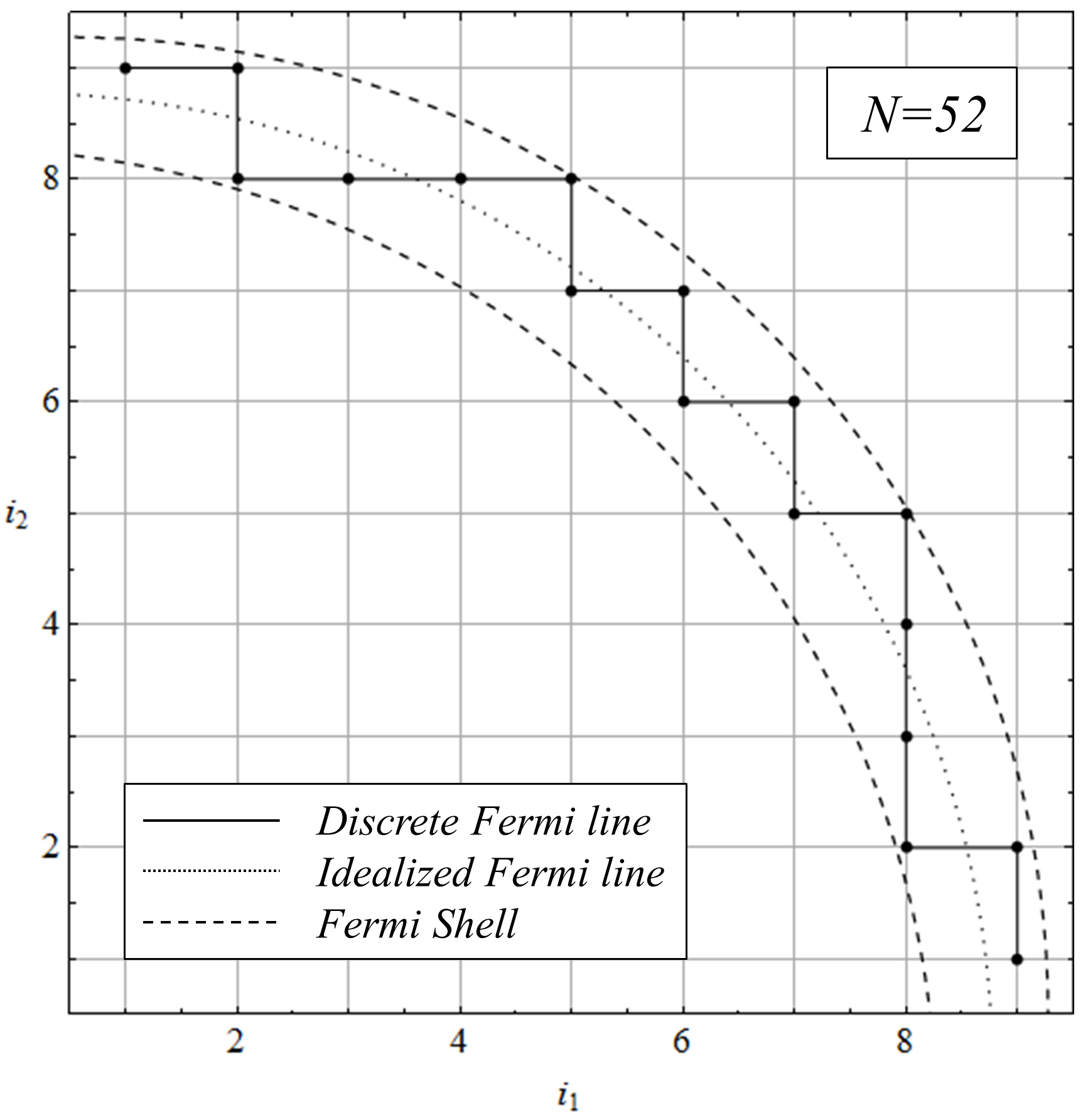}
\caption{Idealized and discrete Fermi lines and Fermi shell for 2D Fermi gas with $\alpha_1=3, \alpha_2=3, \alpha_3=40$ and $N=52$.}
\label{fig:pic3}
\end{figure}

When $N$ or $\Lambda$ increases, Fermi line extends and Fermi shell starts to enclose more and more states. Even if the number of states in Fermi shell increases, FWHM of Fermi line decreases and most of the contributions of states in Fermi shell to heat capacity nearly vanish. Hence, despite the oscillations, when $N\rightarrow\infty$ or $\alpha\rightarrow\infty$, $C_V$ goes to zero.

\begin{figure}[b]
\centering
\includegraphics[width=0.47\textwidth]{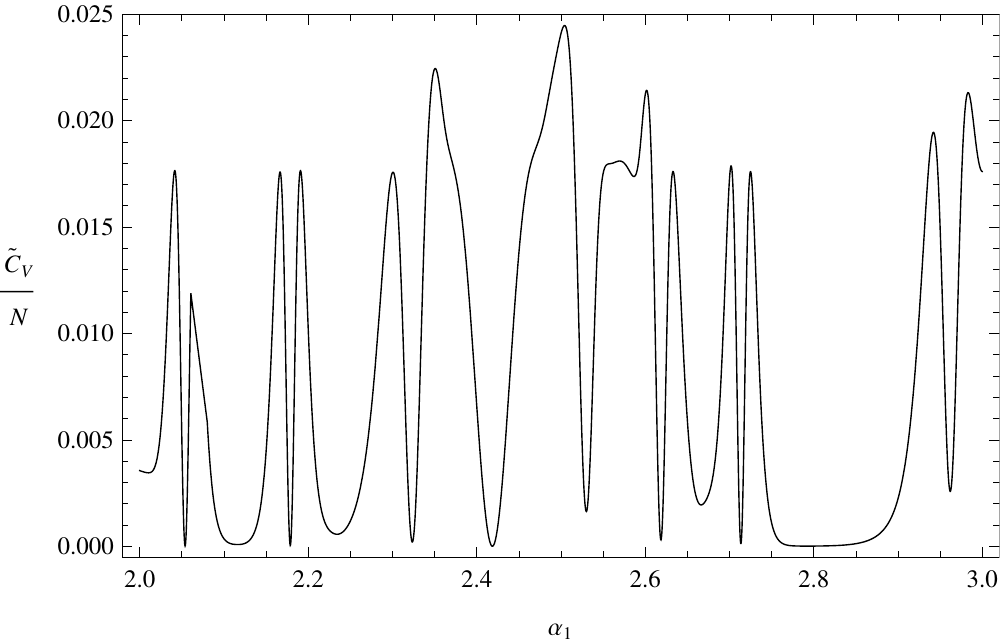}
\caption{Dimensionless heat capacity per particle vs confinement parameter $\alpha_{1}$ for 2D Fermi gas with $\alpha_2=3$, $\alpha_3=40$ and $N=50$.}
\label{fig:pic3}
\end{figure}

In Figure 8, unusual effect of confinement on heat capacity is shown. When $\alpha$ (confinement) increases, quasi-irregular, non-periodic oscillations appear on specific heat capacity of 2D Fermi gases. This behavior again rises due to discrete nature of Fermi line. When the domain size is anisotropically changed, distribution and contributions of states in Fermi shell also change and cause oscillations seen in Fig. 8. On the other hand, when $\alpha$ values for all directions are increased at the same rate, oscillations almost disappear and heat capacity changes smoothly. The reason is that the distance of state points to the idealized Fermi line does not change while the sharpness of the variance function, Eq.(15), changes at the same rate for all the states in Fermi shell.
\subsubsection{Discrete Fermi Surface and Heat Capacity Oscillations in Confined 3D Fermi Gas}
In 3D Fermi gas, contribution to heat capacity comes from the states on Fermi surface. Similar to 2D case, idealized Fermi surface (solid dark gray surface) as well as its $\pm 1/2$ neighborhoods (solid light gray surfaces) and the exact or discrete Fermi surface (meshed surface) are shown in Fig. 9. Again, due to the nature of Fermi-Dirac variance function, contribution to heat capacity comes from the states within Fermi shell. Therefore, instead of making summations over $\left\{i_1, i_2, i_3\right\}$ from 1 to $\infty$, it is enough to sum over $i_1$ from 1 to $i_{1max}=\sqrt{\Lambda}/\alpha_1 + 1/2$, $i_2$ from $i_{2min}=\mbox{Ceiling}(\sqrt{\Lambda-[\alpha_1(i_1+1/2)]^2-[\alpha_3(i_3+1/2)]^2}/\alpha_2 - \frac{1}{2})$ to $i_{2max}=\mbox{Floor}(\sqrt{\Lambda-[\alpha_1(i_1-1/2)]^2-[\alpha_3(i_3-1/2)]^2} \\* /\alpha_2 + 1/2)$ and $i_3$ from 1 to $i_{3max}=\sqrt{\Lambda}/\alpha_3 + 1/2$.

\begin{figure}[H]
\centering
\includegraphics[width=0.48\textwidth]{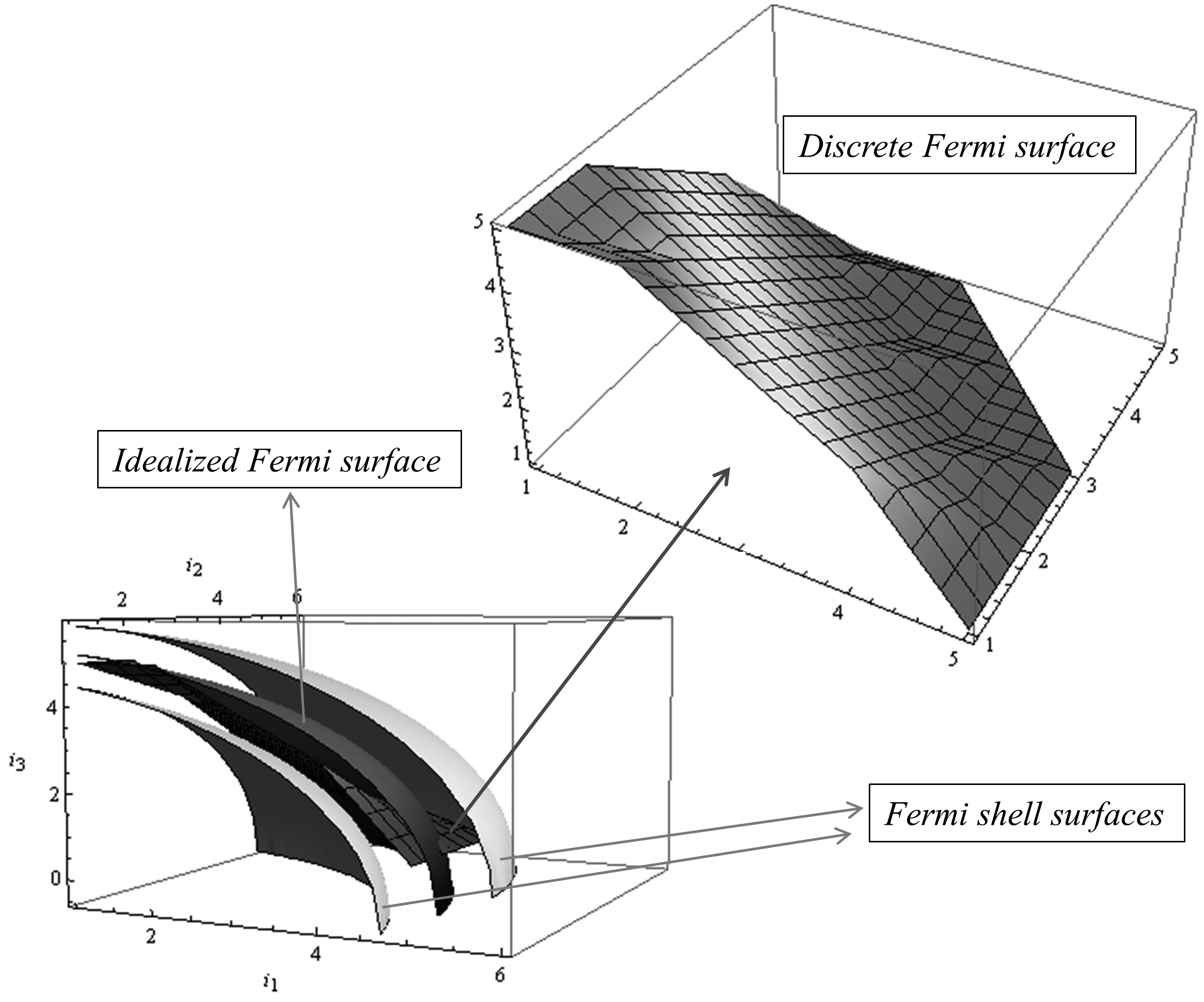}
\caption{Idealized and discrete Fermi surfaces and Fermi shell for 3D Fermi gas with $\alpha_1=3, \alpha_2=3, \alpha_3=3$ and $N=50$.}
\label{fig:pic3}
\end{figure}

\begin{figure}[b]
\centering
\includegraphics[width=0.47\textwidth]{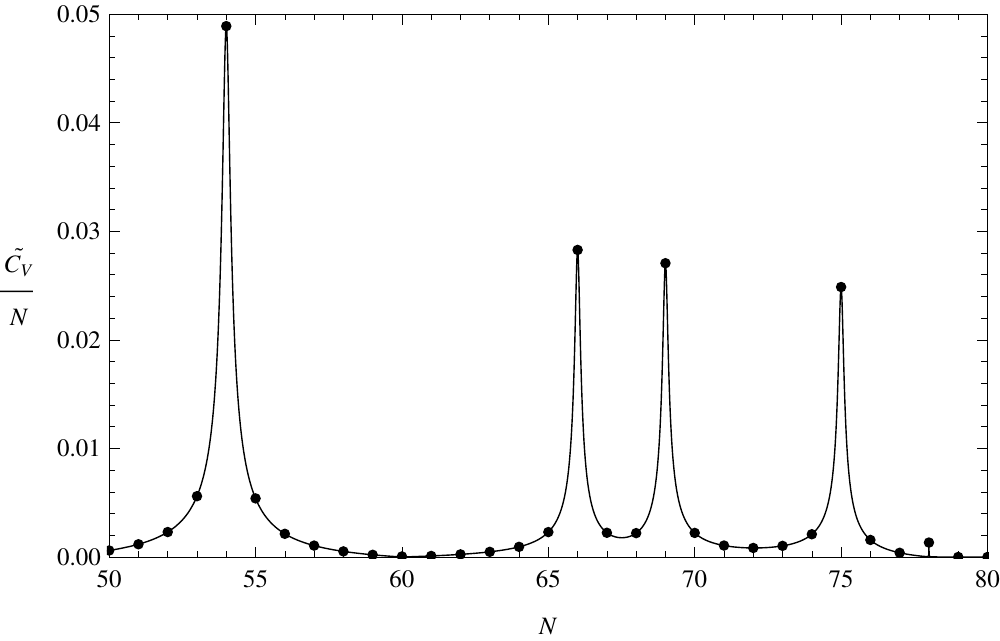}
\caption{Dimensionless heat capacity per particle vs number of particles for 3D Fermi gas with $\alpha_1=3, \alpha_2=3, \alpha_3=3$.}
\label{fig:pic6}
\end{figure}

Similar to 2D case, there are significant sharp peaks in dimensionless heat capacity per particle vs number of particles graph, Figure 10.

Anomalies seen in the variation of heat capacity per particle with number of particles can be interpreted based on all the explanations that are used for 2D case.

If one or two directions are confined, non-periodic oscillations in specific heat capacity appear. In Fig. 11, oscillations in heat capacity per particle due to the variation of confinement rate in the first direction ($\alpha_{1}$) are shown for a 3D Fermi gas. On the other hand, similar to 2D case, when the sizes of an isometric system is reduced in all directions at the same rate, $C_{V}-\alpha$ relation becomes a smooth decaying function instead of frequently oscillatory one.

\begin{figure}[t]
\centering
\includegraphics[width=0.47\textwidth]{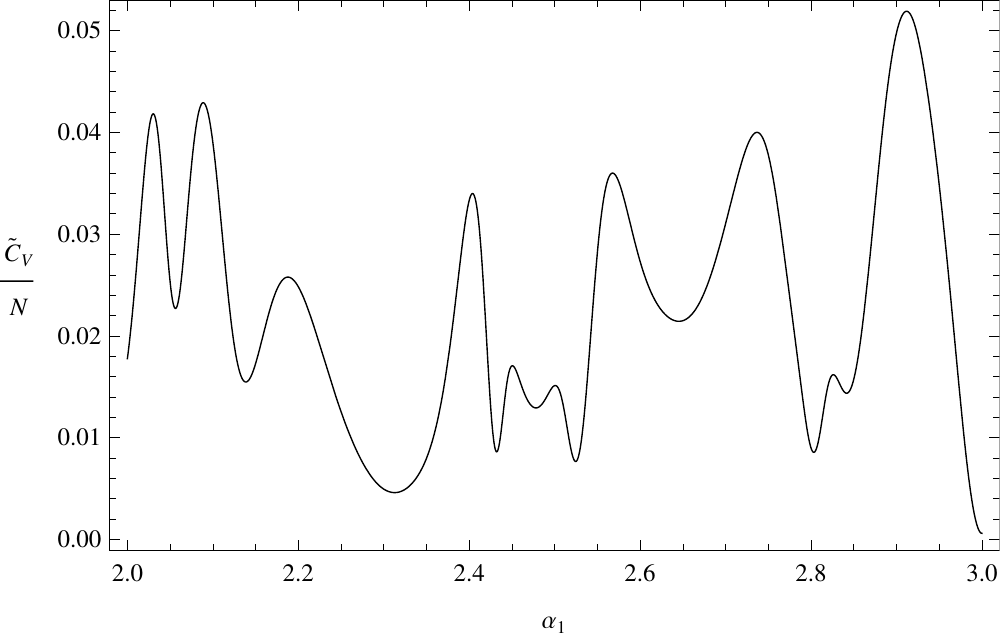}
\caption{Dimensionless heat capacity per particle vs confinement parameter $\alpha_{1}$ for 3D Fermi gas with $\alpha_2=3$, $\alpha_3=3$ and $N=50$.}
\label{fig:pic7}
\end{figure}
\section{Possibility of Experimental Verification and Excess Thermal Energy Storage at Nano Scale}

Advances in nano technologies may provide an experimental verification for heat capacity anomalies (peaks and oscillations) in Fermi gases at nano scale. Consider an isometric 3D InSb (Indium antimonide) nanobox with domain sizes $L_1=L_2=L_3=10$ nm and conduction electron density $9.35\mbox{x}10^{25}$ $\mbox{m}^{-3}$ at $T=5$K \cite{21}. By changing the domain in the first direction from $10$ nm to $12$ nm, while keeping the conduction electron density between the range of $9.31\mbox{x}10^{25}$ $\mbox{m}^{-3}$ and $9.40\mbox{x}10^{25}$ $\mbox{m}^{-3}$ (by adjusting the number of electrons during the size reduction), one may observe the oscillations in the electronic contribution to the specific heat. For this system, particle number changes between 94 and 112 which is far from fluctuations. Due to possible experimental or computational errors, there may be inconsistencies in the comparison of experimental and theoretical results. To avoid this, one can repeat the experiment for the $T=7$K and compare the ratio of the results at two temperatures with our theoretical estimate.

\begin{figure}[h]
\centering
\includegraphics[width=0.48\textwidth]{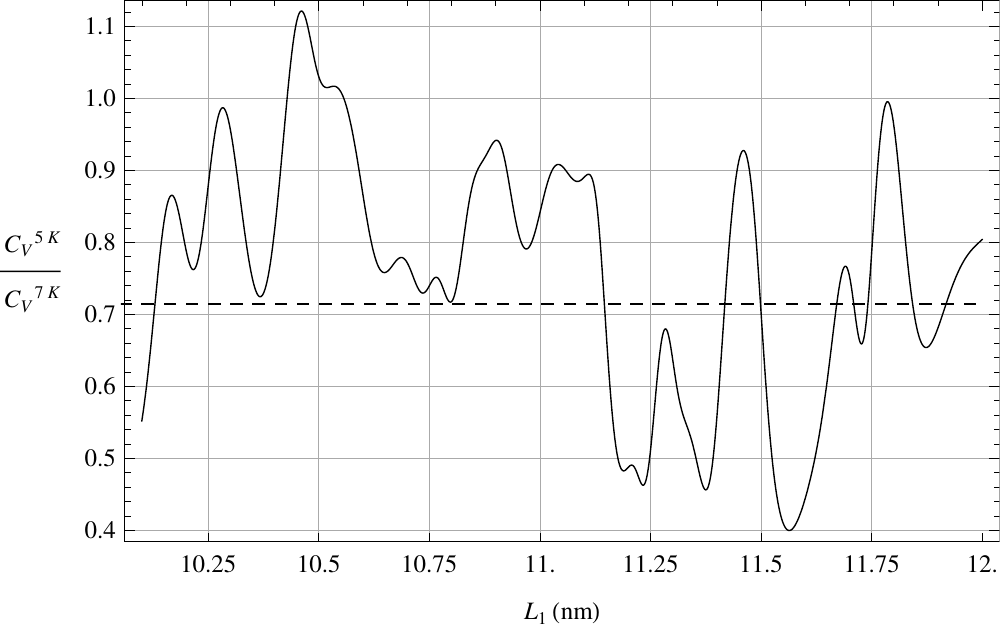}
\caption{Ratio of heat capacity at $T=5$K and $T=7$K varying with domain size in the first direction, $L_{1}\mbox{(nm)}$, for 3D electron gas confined in InSb with $N=\left\{95,...,112\right\}$, $L_2=10$ nm, $L_3=10$ nm and conduction electron density around $n=9.35\mbox{x}10^{25}$ $\mbox{m}^{-3}$.}
\label{fig:pic7}
\end{figure}

In Fig 12, solid curve is obtained by using the exact heat capacity expression in Eq. (14), while the dashed line is achieved by using the heat capacity per particle equation in continuum limit (without size effects), which is expressed as $\widetilde{C}_V/N=\pi^2/2\Lambda$. It is seen that there are considerable and probably measurable deviations from the classical behavior. When domain size becomes comparable to $L_c$, the wave functions of individual particles and the whole system is greatly affected by the boundaries of the domain. When the system is too small, any tiny changes in size greatly affects the system. So, specific heat capacity of Fermi gases at nano scale is strongly depend on anisometric size changes. This may lead us to make excess thermal energy storage devices at nano scale. By making small changes in domain size for one or two direction properly, one can increase or decrease the heat capacity of the substance. Thus, by adjusting domain sizes, high capacity thermal energy storage systems can be developed in nano scale.
\section{Conclusion}
In this paper, it is shown that thermodynamic properties of an ideal Fermi gas can have discrete nature in nano scale. Appearance of discrete nature is the combined consequence of internal features of the system such as Pauli exclusion principle and the quadratic energy-momentum dispersion relation, and the external parameters like quantum degeneracy and quantum confinement. Therefore, discrete nature of thermodynamic properties disappears in Bose and Maxwell gases or Fermi gas in thermodynamic limit (infinite domain size). It is seen that, as macroscopic properties of system, chemical potential and internal energy per particle can have discrete nature. QSE due to the strong confinement reveal themselves as stepwise behaviors in chemical potential and internal energy whereas peakwise and oscillatory behavior in heat capacity. The contributions from the states within Fermi shell centralized by idealized Fermi point, line and surface provide an explanation for unusual behaviors of heat capacity in 1D, 2D and 3D systems respectively. It is also shown that change in domain sizes strongly affects the behavior of heat capacity of Fermi gas at nano scale.

In nano systems with small number of particles, there is a need for the comparison of the order of fluctuations in particle number and energy with the order of discrete corrections based on QSE. The comparison of the order of fluctuations and QSE corrections has been discussed in literature \cite{4}. For the cases considered here, it is seen that order of fluctuations in thermodynamic quantities are much less than the order of quantum (discrete) corrections. Therefore, it may be possible to experimentally verify these results by measuring the heat capacities of nano scale structures at low temperatures since the electronic contribution becomes more significant than the lattice one at low temperature conditions. Verification of discrete nature can also be done by measuring Fermi surface of a strongly confined structures \cite{22}. Possibility of creating new technologies like nano scale thermal energy storage devices based on these effects is proposed.

\end{document}